\documentclass[preprint,aps,pra,superscriptaddress]{revtex4-1}

\usepackage{graphicx,siunitx,amsmath,physics,mathtools,bbold,xcolor}
\usepackage[hidelinks]{hyperref}
\usepackage[utf8x]{inputenc}

\begin{document}

\title{1GHz clocked electrically driven source of single entangled telecom photon pairs}

\author{G. Shooter}
\affiliation{Toshiba Research Europe Limited, Cambridge Research Laboratory, 208 Cambridge Science Park, Milton Road, Cambridge, CB4 0GZ, United Kingdom}%
\affiliation{Cavendish Laboratory, University of Cambridge, J.J. Thomson Avenue, Cambridge, CB3 0HE, United Kingdom}%

\author{Z. Xiang}
\affiliation{Toshiba Research Europe Limited, Cambridge Research Laboratory, 208 Cambridge Science Park, Milton Road, Cambridge, CB4 0GZ, United Kingdom}%
\affiliation{Cavendish Laboratory, University of Cambridge, J.J. Thomson Avenue, Cambridge, CB3 0HE, United Kingdom}%

\author{J.R.A. M\"uller}
\affiliation{Toshiba Research Europe Limited, Cambridge Research Laboratory, 208 Cambridge Science Park, Milton Road, Cambridge, CB4 0GZ, United Kingdom}%
\affiliation{Department of Physics and Astronomy, University of Sheffield, Hounsfield Road, Sheffield S3 7RH, UK}%

\author{J. Skiba-Szymanska}
\affiliation{Toshiba Research Europe Limited, Cambridge Research Laboratory, 208 Cambridge Science Park, Milton Road, Cambridge, CB4 0GZ, United Kingdom}%

\author{J. Huwer}
\affiliation{Toshiba Research Europe Limited, Cambridge Research Laboratory, 208 Cambridge Science Park, Milton Road, Cambridge, CB4 0GZ, United Kingdom}%

\author{J. Griffiths}
\affiliation{Cavendish Laboratory, University of Cambridge, J.J. Thomson Avenue, Cambridge, CB3 0HE, United Kingdom}%

\author{T. Mitchell}
\affiliation{Cavendish Laboratory, University of Cambridge, J.J. Thomson Avenue, Cambridge, CB3 0HE, United Kingdom}%

\author{M. Anderson}
\affiliation{Toshiba Research Europe Limited, Cambridge Research Laboratory, 208 Cambridge Science Park, Milton Road, Cambridge, CB4 0GZ, United Kingdom}%
\affiliation{Cavendish Laboratory, University of Cambridge, J.J. Thomson Avenue, Cambridge, CB3 0HE, United Kingdom}%

\author{T. M\"uller}
\affiliation{Toshiba Research Europe Limited, Cambridge Research Laboratory, 208 Cambridge Science Park, Milton Road, Cambridge, CB4 0GZ, United Kingdom}%

\author{A.B. Krysa}
\affiliation{EPSRC National Epitaxy Facility, Department of Electronic \& Electrical Engineering, The University of Sheffield, 3 Solly Street, Sheffield, S1 4DE }%

\author{R. M. Stevenson}
\affiliation{Toshiba Research Europe Limited, Cambridge Research Laboratory, 208 Cambridge Science Park, Milton Road, Cambridge, CB4 0GZ, United Kingdom}%

\author{J. Heffernan}
\affiliation{EPSRC National Epitaxy Facility, Department of Electronic \& Electrical Engineering, The University of Sheffield, 3 Solly Street, Sheffield, S1 4DE }%

\author{D. A. Ritchie}
\affiliation{Cavendish Laboratory, University of Cambridge, J.J. Thomson Avenue, Cambridge, CB3 0HE, United Kingdom}%

\author{A. J. Shields}
\affiliation{Toshiba Research Europe Limited, Cambridge Research Laboratory, 208 Cambridge Science Park, Milton Road, Cambridge, CB4 0GZ, United Kingdom}%

\date{23/11/20}

\begin{abstract}
Quantum networks are essential for realising distributed quantum computation and quantum communication. Entangled photons are a key resource, with applications such as quantum key distribution, quantum relays, and quantum repeaters. All components integrated in a quantum network must be synchronised and therefore comply with a certain clock frequency. In quantum key distribution, the most mature technology, clock rates have reached and exceeded 1GHz. Here we show the first electrically pulsed sub-Poissonian entangled photon source compatible with existing fiber networks operating at this clock rate. The entangled LED is based on InAs/InP quantum dots emitting in the main telecom window, with a multi-photon probability of less than 10\% per emission cycle and a maximum entanglement fidelity of 89\%. We use this device to demonstrate GHz clocked distribution of entangled qubits over an installed fiber network between two points 4.6km apart.
\end{abstract}

\maketitle

© 2020 Optica Publishing Group. Users may use, reuse, and build upon \href{https://www.osapublishing.org/oe/fulltext.cfm?uri=oe-28-24-36838&id=442793}{\textcolor{blue}{the article}}, or use the article for text or data mining, so long as such uses are for non-commercial purposes and appropriate attribution is maintained. All other rights are reserved.

\section*{Introduction}

For implementation of various kinds of advanced quantum network schemes \cite{Ekert.1991,Briegel.1998,Jacobs.2002,Riedmatten.2004}, entanglement must be distributed between nodes \cite{Komar.2014,sun.2016,valivarthi.2016,Wengerowsky.2019}. The most widely used sources are currently based on spontaneous non-linear processes \cite{Kwiat.1995,Li.2005}, though the efficiency of these sources is intrinsically limited if multi-photon emission is to be minimised. This limit does not apply to entangled photon sources with sub-Poissonian statistics, such as semiconductor quantum dots (QD)s \cite{Benson.2000,Michler.2000}, with the prospect of deterministic entangled pair generation \cite{liu.2019}.

For an entangled photon source to be embedded in a quantum network, it must further conform to the basic requirements of operating clock rate and wavelength. State-of-the art quantum key distribution (QKD) systems operate at clock frequencies of 1GHz and above \cite{Yuan.2008,boaron.2018}, with photons in the telecom C-band most suitable for distribution over standard optical fibers.

Epitaxially grown semiconductor QDs can be readily incorporated into PIN diode structures, enabling the fabrication of light sources using standard semiconductor processing techniques \cite{Benson.2000,Michler.2000b}. As QDs embedded within diodes can be electrically excited \cite{Yuan.2002}, it is possible to create entangled photon sources that can be conveniently operated similar to other standard light sources, such as telecom laser diodes. InAs/InP QDs emit in the lowest-loss silica fiber window \cite{SkibaSzymanska.2017}, which makes them prime candidates for transmission over standard fiber networks.

Entangled LED (ELED) telecom C-band sources have been demonstrated with DC excitation  \cite{Muller.2018}.  Pulsed single \cite{Michler.2000b,Santori.2001,Hargart.2013} and entangled \cite{Benson.2000,Yuan.2002,Stevenson.2006,zhang.2015,Varnava.2016,Muller.2020} photon sources based on semiconductor QDs have been developed but are either at short wavelengths, and therefore incompatible with existing fiber networks, or only operate at repetition rates too slow for current quantum network applications. Entanglement distribution experiments over installed networks have used low repetition rates \cite{sun.2016,valivarthi.2016,Wengerowsky.2019} while GHz clock rates, necessary for synchronisation with high clock rate QKD systems, have only been demonstrated with nonlinear sources over long fiber in a laboratory \cite{Honjo.2007,Inagaki.2013}. In this work, we show the distribution of  entangled qubits from a 1GHz driven sub-Poissonian source over an installed standard telecom network.

\section*{GHz clocked single photon source}

\begin{figure}[ht]%
\centering
\includegraphics[width=15cm]{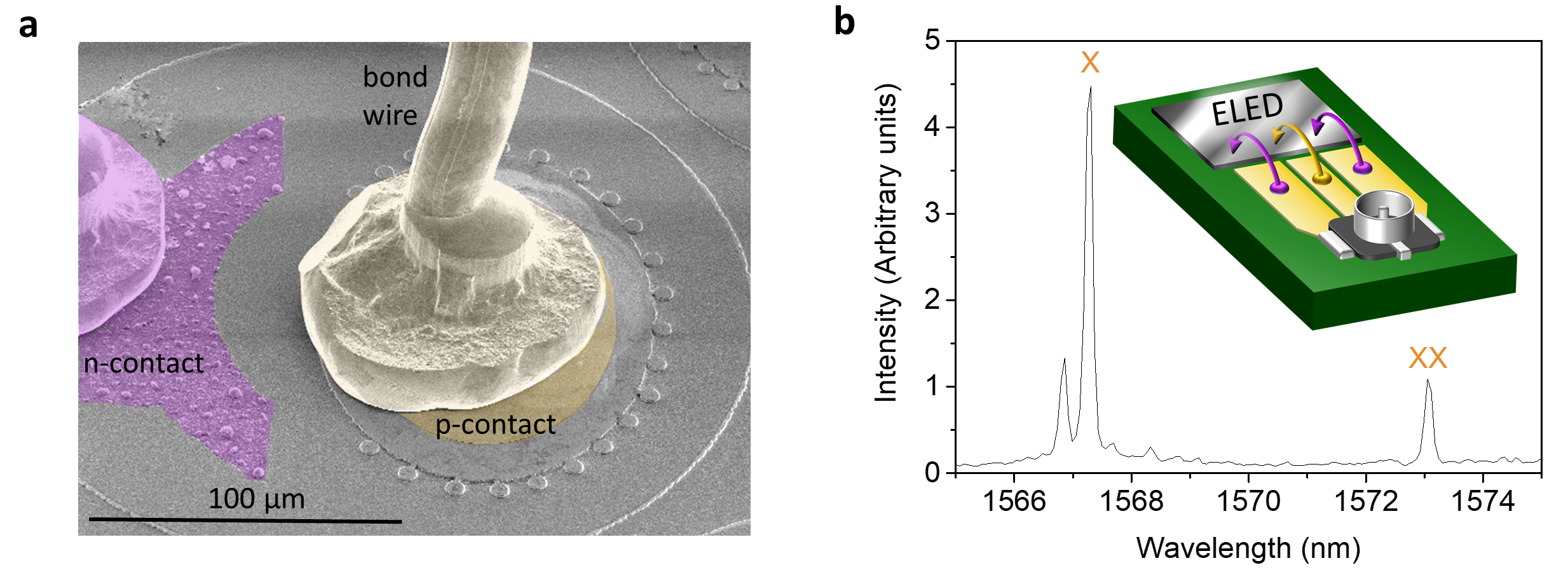}
   \caption{A quantum dot LED for 1GHz pulsed electrical excitation. (a) A false colour scanning electron microscope image of a device; an oval mesa with connected pillars, electrically isolated from the surrounding wafer. (b) An electroluminescence (EL) spectrum of a quantum dot showing the exciton (X) and biexciton (XX) emission lines. The inset shows an illustration of the radio frequency compatible packaging; the ELED is wire bonded to impedance matched tracks (p-type and n-type bond are shown in yellow and purple respectively) which finish at a low-profile micro-coaxial connector.}
    \label{fig:DEVICE}
\end{figure}

The fabrication of ELED devices used in this work was developed to be simple, with only two etch steps and two metal depositions.  An image of a device is shown in Fig.\ref{fig:DEVICE}(a). This design allows for fast electrical operation at GHz frequencies, with dimensions close to the limit imposed by the size of a bond ball as can be seen in Fig. \ref{fig:DEVICE}(a). The ELED shows good electrical performance as a diode, with the resistance reaching 50$\Omega$ beyond the turn-on voltage. The wafer structure, described in the Supplemental Material, is designed for QD emission in the telecom C-band. The emission spectrum shown in Fig.\ref{fig:DEVICE}(b) comes from a QD that is located within a 5$\mu$m connected pillar, as can be seen in Fig.\ref{fig:DEVICE}(a), and so can be readily relocated. The device was mounted onto the centre of a radio-frequency compatible FR4 packaging with conductive paint before wire bonding to a Au layer on one end of 50$\Omega$ impedance matched tracks, ending with a low-profile micro-coaxial connector as illustrated in the inset of Fig.\ref{fig:DEVICE}(b).

In QDs, single photons are emitted via the radiative recombination of confined electron-hole (e-h) pairs \cite{Imamoglu.1994}. Entangled photon pairs are emitted via the biexciton cascade \cite{Benson.2000} where a QD initialised in the doubly excited biexciton (XX) state decays to the singly excited exciton (X) state via emission of the first photon. This state subsequently decays via emission of a second photon, leaving the QD in the ground state. Due to conservation of angular momentum, the two emitted photons are maximally entangled in their polarization.

\begin{figure}[ht]
    \centering
    \includegraphics[width=15cm,keepaspectratio]{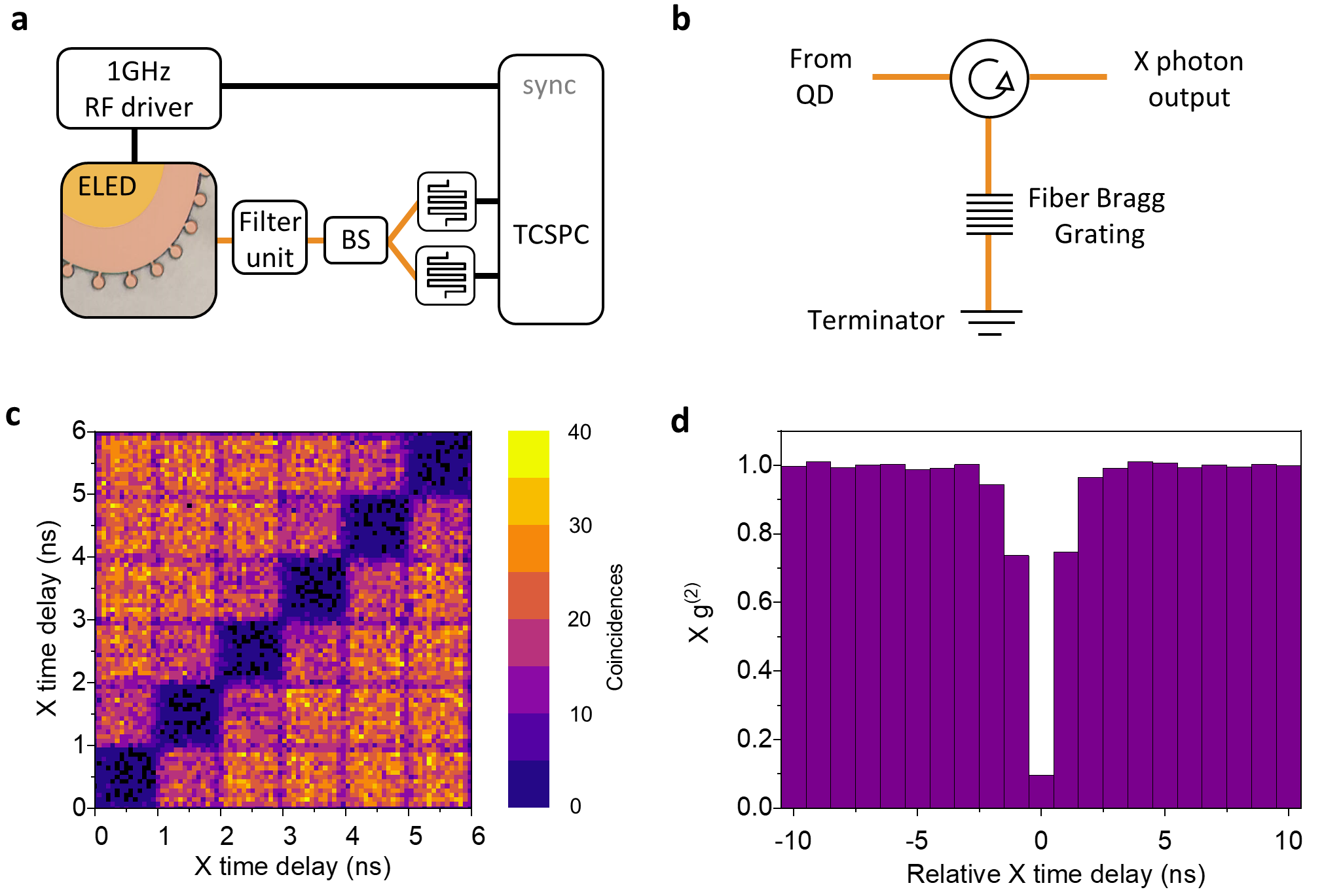}
    \caption{Measurement of 1 GHz pulsed single photon emission. (a) Experimental setup for measuring the second order autocorrelation (g$^{(2)}$). Light from the ELED passes through a fiber-based spectral filter unit followed by a fiber 50:50 beam splitter (BS), with each output detected using superconducting nanowire single photon detectors and a time correlated single photon counter (TCSPC). (b) A diagram of the filter unit, with a circulator followed by a fiber bragg grating which reflects the X photons to the fiber output. (c) Time resolved coincidences of 1GHz clocked X photons for 6 consecutive emission cycles (72ps time bins) as a function of detection time with respect to the clock. Time bins with no photon coincidences are black. (d) A histogram of normalised coincidences with 1ns bins for 21 relative time delays.}
    \label{fig:g2}
\end{figure}

To assess the sub-Poissonian photon emission from the ELED, we measure the second order autocorrelation function (g$^{(2)}$) of X photons as shown in Fig.\ref{fig:g2}(a). This measurement requires isolation of the X spectral line as in Fig.\ref{fig:DEVICE}(b). Since the QD emits at telecom wavelengths, a compact spectral wavelength filtering unit can be used that is based on an optical add-drop multiplexer as shown in Fig.\ref{fig:g2}(b) (FWHM $<$0.25nm, 1.31dB loss), a common component in classical telecommunication technology.

The ELED was electrically driven with $\sim$130ps FWHM pulses with a high of 1.5V and a low of 0.3V; the arrival times of photons at each detector were recorded with respect to the 1GHz clock of these pulses. Photon correlations acquired over 90 minutes can be seen in Fig.\ref{fig:g2}(c). The 1ns squares along the bottom-left to top-right diagonal contain coincidences of photons emitted within the same excitation cycle. These are strongly reduced compared to the rest of the grid, corresponding to an excellent suppression of multi-photon emission within the same excitation cycle. Coincidences in the 1ns squares adjacent to the bottom-left to top-right diagonal are also reduced, as the cascade is not fully reinitialised each cycle by this electrical driving condition.

Coincidences in Fig.\ref{fig:g2}(c) are also suppressed in a grid pattern with spacing of 1ns. This pattern occurs due to the electrical excitation pulse at the start of each 1ns cycle, when the QD is reinitialised. During this reinitialisation period, the population in the X level is depleted due to excitation to higher energy levels such as the XX. The 1ns squares containing coincidences of photons emitted in different excitation cycles appear to have an almost flat distribution. This is due to the long natural lifetime of the X state of 1.9ns and the dynamics involved in populating the X state via decay from the XX state, which has a lifetime of 0.5ns.

Photon coincidences in each 1ns square were then normalised using coincidences in cycles with completely uncorrelated detection events. Fig.\ref{fig:g2}(d) shows that the g$^{(2)}$ for X photons emitted in the same 1ns cycle is 0.097$\pm$0.002 without application of any temporal post-selection. This is far below the classical limit, proving strongly sub-Poissonian emission. The g$^{(2)}$(0) is limited by the non-resonant excitation scheme used here, likely due to interactions with the charge environment.

\clearpage

\section*{GHz clocked entanglement}

\begin{figure}[ht]
    \centering
    \includegraphics[width=14cm,keepaspectratio]{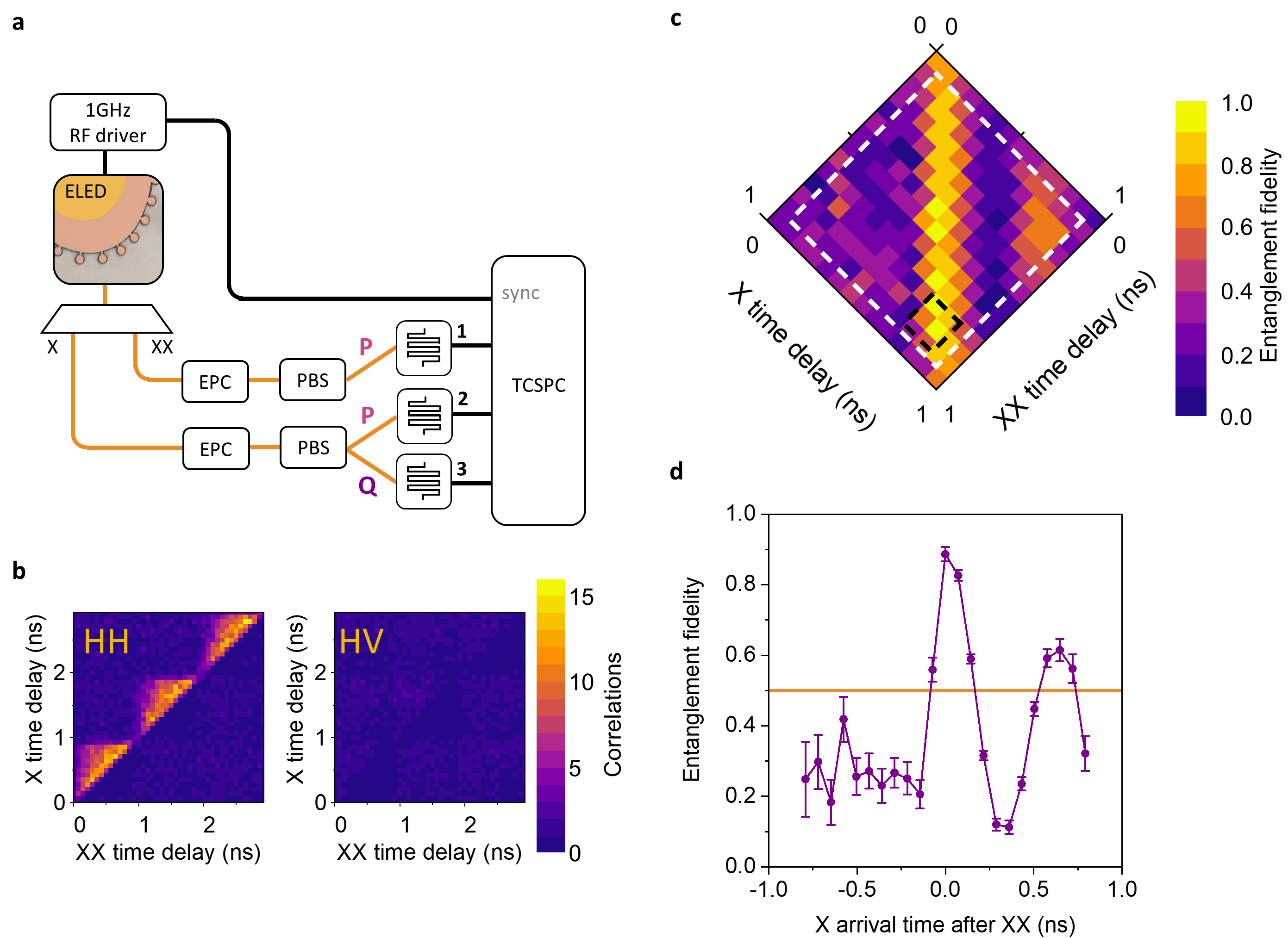}
    \caption{Measurement of entanglement. (a) The experimental setup with a free-space spectral filter separating XX and X photons.  Electronic polarization controllers (EPC)s followed by polarizing beam splitters (PBS)s set the detection polarization basis, with time correlated single photon counters (TCSPC)s recording photon arrivals at superconducting nanowire single photon detectors. (b) Photon correlations of horizontally (H) and vertically (V) polarized biexciton (XX) and exciton (X) photons, with co- and cross-polarized detection (left and right panels respectively) as a function of XX and X delay with respect to the 1GHz clock (72ps time bins). (c) The fidelity to the maximally entangled Bell $\phi^{+}$ state with the same XX and X photon time bins as (b) but rotated by 135$^{\circ}$. Photon arrival times gated to the central 0.864ns of each 1ns are shown by the dashed white square. A gate of 0.168ns is shown by the dashed black square. (d) The fidelity to the maximally entangled Bell $\phi^{+}$ state as a function of relative delays between XX and X photons (72ps time bins). The horizontal orange line shows the classical limit of the entanglement fidelity.}
    \label{fig:entResults}
\end{figure}

As entangled photons are critical for quantum network applications, we now show the generation of 1GHz clocked entangled photon pairs from our ELED. The device was driven similarly to before, using pulses with a high of 1.5V and a low of 0.5V. The XX and X photons were separated with a spectral filter and each were detected with a polarization analyser comprising of an electronic polarisation controller (EPC) and a polarising beam splitter (PBS) followed by superconducting nanowire single photon detectors (SNSPD)s as in Fig.\ref{fig:entResults}(a). When detecting photons in a polarization basis PQ, P polarized XX photons were measured at detector 1 in Fig.\ref{fig:entResults}(a), with X photons of P and Q  polarizations measured at detectors 2 and 3 respectively. As only one output of the XX PBS was sent to a detector, only half of all possible photon coincidences were recorded. Photon arrival times for each detector with respect to the 1GHz clock signal divided by 64 were recorded with a time correlated single photon counter (TCSPC), with photon correlations and entanglement fidelity evaluated in postprocessing.

Photon pair correlations in the horizontal-vertical (HV) basis covering three consecutive excitation cycles can be seen in Fig.\ref{fig:entResults}(b). Within a 1ns cycle both co (HH) and cross (HV) -polarized photon correlations are suppressed for time bins corresponding to the arrival of an X photon before a XX photon, due to the cascaded emission of the two photons of a pair. For time bins corresponding to the arrival of a XX photon before an X photon, only correlations for co-polarized photons are observed, as expected for the maximally entangled Bell $\phi^{+}$ state. Importantly, correlations do not extend beyond each 1ns excitation cycle. The cascade can be seen to decay to the uncorrelated level within 1 excitation cycle (Supplemental Material, Fig.\ref*{fig:coincs}), verifying a clean reinitialisation of photon pair emission at a 1GHz rate.

The resulting fidelity to the maximally entangled Bell $\phi^{+}$ state, calculated as explained in the Supplemental Material, is shown in Fig.\ref{fig:entResults}(c). For most of the grid in Fig.\ref{fig:entResults}(c), the entanglement fidelity is $\sim$0.25, corresponding to completely random polarization correlations. For XX and X photons from the same 1ns cycle, the entanglement fidelity rises above the classical limit of 0.5 for X photons arriving after XX photons, before decaying with oscillations due to the fine structure splitting of the QD of 6.0$\mu$eV \cite{stevenson.2008}. These oscillations are caused by a quantum beat in the superposition bases, as can be seen in the Supplemental Material, Fig.\ref*{fig:corrLab}.

Each vertical column of time bins in Fig.\ref{fig:entResults}(c) contains photon coincidences with the same relative XX-X time delay, but different arrival times within the 1ns emission cycles.  One can see that the entanglement fidelity drops for time bins at the start and end of the 1ns cycles due to reinitialisation of the emission ($\sim$130ps). Therefore, to give an idea of the highest possible value, XX and X photon arrival times were additionally gated to 0.864ns around the center of 1ns cycles (shown as a white dashed box in Fig.\ref{fig:entResults}(c)). The average of each column is plotted in Fig.\ref{fig:entResults}(d), where one can again observe the time dependent oscillation of the fidelity due to the finite fine structure splitting of the QD. The resulting maximum fidelity to the Bell $\phi^{+}$ state is 0.89$\pm$0.02 with comparable correlation contrasts in the three principal polarisation bases (see the Supplemental Material for further information). However, this value corresponds to a bin size of 72ps, which is not compatible with post-selection free detection schemes.

Detectors used in state-of-the-art QKD systems operating at 1GHz clock rates have typical detection gate widths of $<$170ps \cite{Yuan.2007,Yuan.2008}. To assess the performance of the pulsed ELED with these non-research grade detectors, we position a single 168ps integration window to give maximum entanglement fidelity, shown as a black dashed box in Fig.\ref{fig:entResults}(c). This results in a fidelity of 0.86$\pm$0.03, in the regime compatible with error correction in quantum key distribution applications \cite{Chau.2002}. The drop in fidelity when increasing the time window size is within the errors, showing that the QD FSS is not limiting the fidelity achievable with typical gated detectors. Analysing the X autocorrelation data from Fig.\ref{fig:g2} in a similar fashion gives a g$^{(2)}$ of 0.04$\pm$0.01.

However, only 3.4$\%$ of the detected photon pairs originating from the same excitation cycle arrive within this 168ps time window. For future compatibility of deterministic GHz clocked entangled photon pair sources with gated detectors for post-selection-free operation, high source efficiencies are crucial. In addition, XX and X lifetimes similar to the detector gate width are necessary to increase the number of photon pairs arriving within the active gate window of the detectors. This could be achieved via Purcell enhancement, which reduces XX and X lifetimes, for example with micropillar designs \cite{Dousse.2010} or circular bragg gratings \cite{liu.2019,Wang.2019}.

Given the GHz clock rate, an overall efficiency of the optical system including detectors of approximately 0.6\% (see the Supplemental Material), and average XX and X photon rates at each detector of 52000 and 83000 counts per second, we estimate an intrinsic efficiency of around 2\% for the ELED to generate a photon per excitation pulse. Efficiencies are currently low for non-resonantly excited telecommunication wavelength QDs, which are still undergoing significant development and are not as well established as short-wavelength InAs/GaAs dots. Telecommunication wavelength QDs are larger than their short-wavelength counterparts, making them more susceptible to fluctuations in the surrounding charge environment. This typically results in the presence of multiple charged states with radiative and non-radiative decay paths. Techniques to enhance emission from the neutral XX and X states rather than charged complexes may increase the photon pair efficiency for ELEDs in the future \cite{young.2007}. Perhaps counter-intuitively, truncating the cascade by reinitialising the QD at a high clock rate does not intrinsically limit the photon generation rates; we have recently shown that photon generation rates can surpass those achievable with DC driving for some pulsed regimes \cite{Muller.2020}.

\section*{Entanglement distribution}

\begin{figure}[ht]
    \centering
    \includegraphics[width=14cm,keepaspectratio]{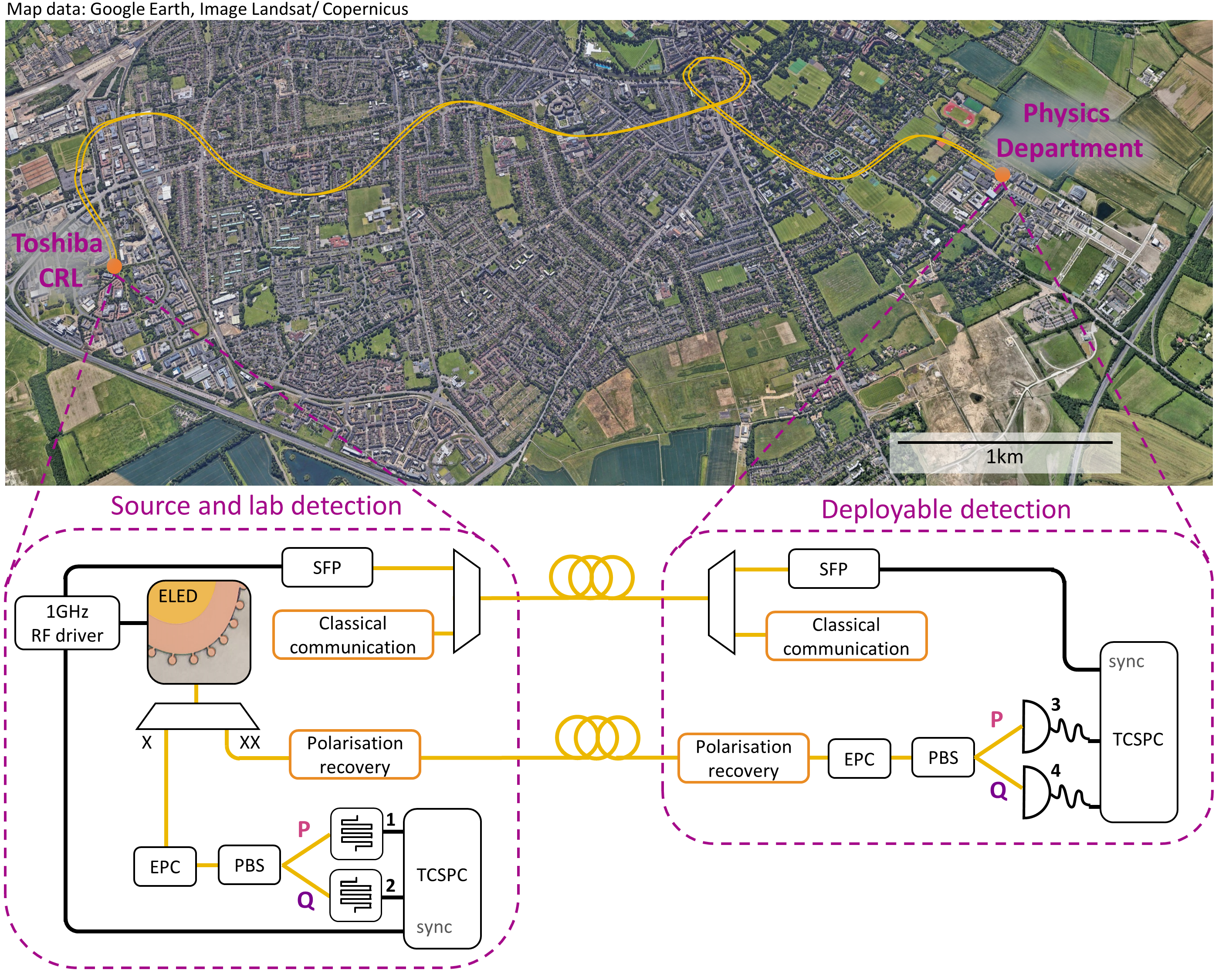}
    \caption{Experimental setup for distribution of 1GHz clocked entangled photon pairs across the city of Cambridge. The entangled photon pairs were generated from an ELED at the Toshiba Cambridge Research Laboratory, with the X photons detected locally using detectors 1 and 2 (SNSPDs) and the XX photons detected at the Physics Department of the University of Cambridge with detectors 3 and 4 (APDs) in a deployed detection system. The detection polarization basis was set by EPCs followed by PBSs, with TCSPCs recording photon arrivals at detectors. When measuring photons in an arbitrary polarization basis PQ, detectors 1 and 3 measured P-polarized X and XX photons respectively and detectors 2 and 4 measured Q-polarized X and XX photons respectively. A polarization recovery system compensated for drifts occurring over the installed fiber. Classical communication for remote control of system components was multiplexed with the reference clock signal from a small form-factor pluggable transceiver (SFP) over a second fiber. }
    \label{fig:entDist}
\end{figure}

To demonstrate network compatibility of the pulsed entangled photon pair source we distributed entanglement over 4.6km between the Toshiba Cambridge Research Laboratory (CRL) and the Physics Department of the University of Cambridge as shown in Fig.\ref{fig:entDist}, using installed network fiber. The source was operated at CRL where X photons were detected, and XX photons were sent to a deployed detection system over 15km of installed fiber with 6dB loss at 1550nm.

The electrical 1GHz clock signal used to drive the ELED was down-sampled to 15.6MHz and converted to an optical signal at 1570nm and multiplexed with 1Gbit/s classical communication data traffic at 1310nm.  The communication channel was required for remote control of the detection system and data acquisition, both classical signals were transmitted over a separate installed fiber. At the other end, both classical signals were demultiplexed, and the clock signal was converted back to an electrical signal to be used as the synchronisation reference in the deployed detection system.

In both locations, photon arrival times in two detector channels were recorded with respect to the reference clock with TCSPCs similar to the   previously discussed measurements in a laboratory. Photon arrival times were measured in the three principal detection bases in sets of 7 minutes. Polarization drifts occurring over the network fiber due to changing environmental conditions were compensated for before each measurement using a similar stabilisation system as in \cite{Xiang.2020}. Photon correlations were evaluated in postprocessing.

\begin{figure}[ht]
    \centering
    \includegraphics[width=14cm,keepaspectratio]{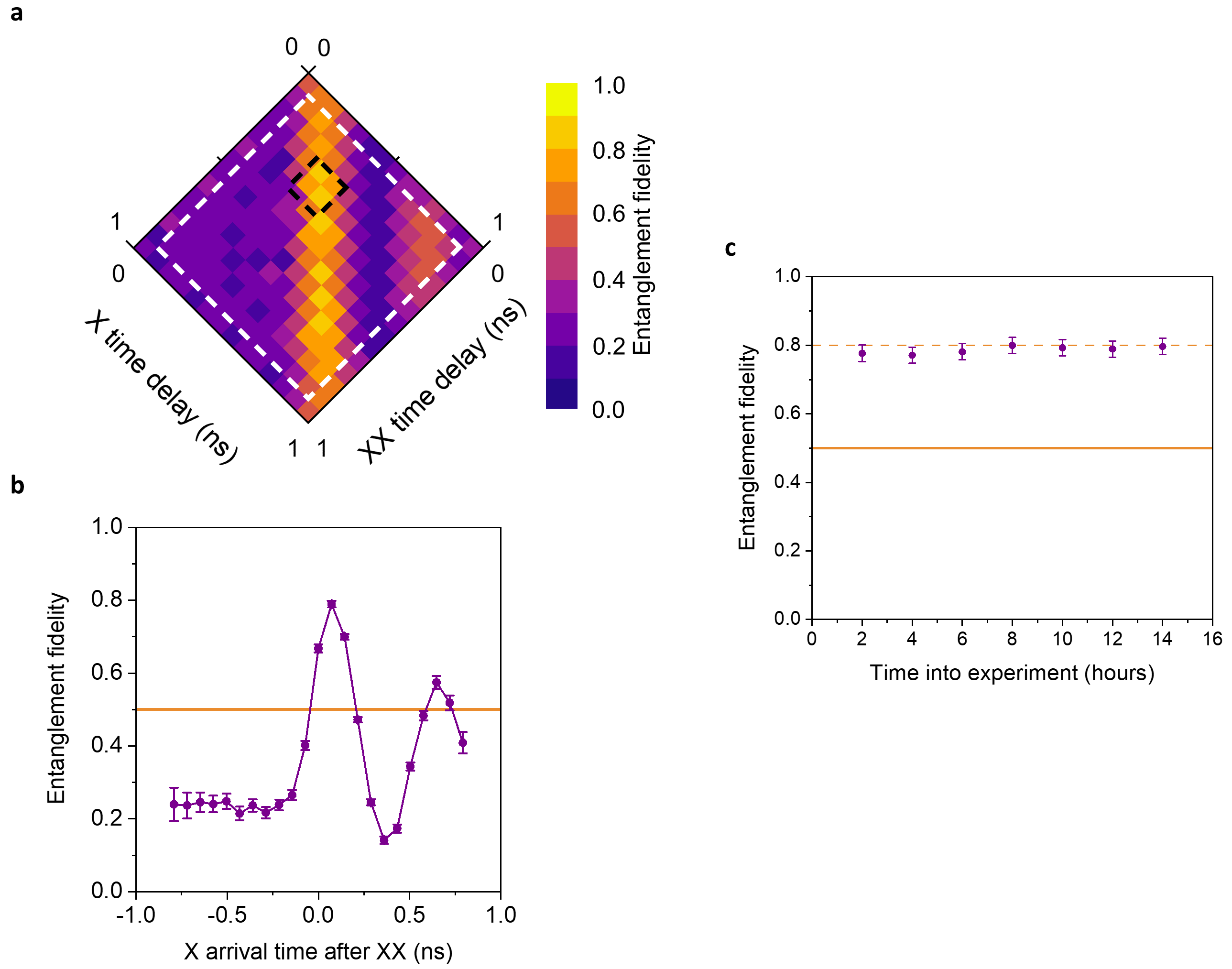}
    \caption{Distribution of entangled photon pairs over 15km of installed fiber. (a) The fidelity to the maximally entangled Bell $\phi^{+}$ state for 14 hours of data acquisition as a function of XX and X delay with respect to the 1GHz clock (72ps time bins) with gating applied similarly to in Fig.\ref{fig:entResults}(c). (b) Fidelity as a function of relative delays between XX and X photons (72ps time bins) . The horizontal orange line shows the classical limit of the entanglement fidelity. (c) The maximum entanglement fidelity from (a) for 2 hour sections throughout the experiment.  The horizontal orange lines show the classical limit of the entanglement fidelity (solid) and a fidelity of 0.8 (dashed).}
    \label{fig:distResult}
\end{figure}

Entangled photon pairs were distributed between East and West Cambridge for 14 consecutive hours of operation. Fig.\ref{fig:distResult} (a) and (b) shows results plotted in a similar way to Fig.\ref{fig:entResults} but for distribution of entanglement rather than a measurement in a laboratory. The maximum fidelity to the Bell $\phi^{+}$ state, analysed on a 72ps grid with the reinitialisation period discarded as for the laboratory measurement, is 0.79$\pm$0.01. Using the timing characteristics of GHz clocked detectors as indicated in Fig.\ref{fig:distResult}(a), the maximum entanglement fidelity is 0.76$\pm$0.01.

The 10\% reduction in the fidelity when transmitting XX photons over the installed fiber is attributed to an increased ratio of background events to XX photon signal at the deployed detectors from $<$2\% to $>$10\%. We further observe a larger drop in polarization correlation contrast for measured superposition bases (diagonal/antidiagonal and right-/left-hand circular, see the Supplemental Material). This most likely results from a larger uncertainty in calibrating the detection bases at the deployed detection system (see the Supplemental Material) which is again caused by a drop in the signal-to-background ratio rather than the performance of the ELED itself.

Fig.\ref{fig:distResult}(c) shows the evolution of the maximum entanglement fidelity for sets of 2 hours of data. It remains around 0.79 for the entire 14 hour experiment, demonstrating the excellent stability of the 1GHz clocked ELED as a source for distributed entangled photon pairs across a real-world fiber network.

\section*{Conclusion}

We have shown an electrically driven 1GHz clocked telecom ELED with strong single photon characteristic, resulting in a two-photon probability of less than 10\% without any temporal post-selection. Using the ELED as a source of 1GHz clocked entangled photons yields a maximum entanglement fidelity of 89\% in a 72ps post selection window. In addition, the device is suitable for operation using standard actively gated GHz clocked detector modules as are used in current QKD systems, with no additional software-based post-selection. However, for real-world applications in quantum communication relying on high entangled photon pair rates, an enhancement of the source brightness is required and the number of photons arriving within the active gate window of such detectors must be significantly increased via the reduction of XX and X lifetimes.

Operation of the device in the lowest-loss telecom window enabled us for the first time to demonstrate the distribution of 1GHz clocked entangled qubits from a sub-Poissonian source on a city scale. The achieved entanglement fidelity of 79\% proves reliability for electrically pulsed semiconductor quantum light sources connected to installed fiber networks. Pulsed operation with a GHz clock frequency opens up the possibility for seamless integration with other quantum network hardware such as QKD systems and efficient time multiplexing with classical communication signals.

A further developed device design with integration of nano-photonic structures \cite{Bockler.2008,Dousse.2010,liu.2019,Wang.2019} to combine high collection efficiency and Purcell enhancement has the potential to provide a viable future workhorse for quantum communication systems, with intrinsic security and no fundamental efficiency limitation as is the case for sources based on weak coherent laser pulses, spontaneous parametric down-conversion or four-wave mixing.

\clearpage

\section*{Acknowledgements}
The authors acknowledge partial financial support from the Engineering and Physical Sciences Research Council and the UK's innovation agency, Innovate UK. Ginny Shooter and Matthew Anderson acknowledge support from Industrial CASE awards funded by the EPSRC and Toshiba Europe Limited. Zi-Heng Xiang acknowledges support from the Cambridge Trust and China Scholarship Council (CSC). Jonathan R.A. M\"uller acknowledges support from the European Union's Horizon 2020 research and innovation programme under the Marie Sk\l{}odowska-Curie grant agreement No 721394.

\clearpage

\appendix

{\large\textbf{Supplemental Material}}

\setcounter{figure}{0}
\makeatletter
\renewcommand{\thefigure}{S\@arabic\c@figure}
\makeatother

\section*{Growth}
The semiconductor wafer was grown by metal organic vapor phase epitaxy on an InP substrate. A bottom distributed Bragg reflector (DBR) comprising 20 layer pairs, each formed of 112nm of (Al$_{0.30}$Ga$_{0.70}$)$_{0.48}$In$_{0.52}$As and 123nm of InP were grown, with the top 3 repeats doped with 2x10$^{18}$cm$^{-3}$ of Si. InAs droplet QDs as in \cite{SkibaSzymanska.2017} were grown in a cavity on a 3$\lambda$/4 intrinsic InP layer followed by a 5$\lambda$/4 InP layer. The top nominal 150nm of the cavity was grown with Zn doping of 2x10$^{18}$cm$^{-3}$ to improve high speed electrical injection at low temperatures. 3 DBR pairs with Zn doping of 2x10$^{18}$cm$^{-3}$ and an InP capping layer completed the p-i-n diode structure.

\section*{Fabrication}
The device in Fig.\ref{fig:DEVICE}(a) was fabricated in 4 steps. To contact the p-layer, CrAu was thermally evaporated onto the wafer surface. The p-type contact was just large enough to fit a bond ball, 80x110$\mu$m. The mesa and the isolated area were each etched using inductively coupled plasma (ICP) with Cl$_{2}$ based process chemistry. 150nm of AuGeNi was evaporated onto the isolated area to contact the n-layer before annealing at 420$^{\circ}$C.

\section*{Characterisation}
The device was cooled to 30K in a He vapour cryostat, with an xyz piezo nano-positioning stage enabling navigation around the device. A fibre-based confocal microscope system with NA 0.68 collected light emitted from the device. QD electroluminescence (EL) spectra were measured by sending this light via a fiber to a spectrometer with an InGaAs array. The fine structure splitting of a QD was measured as in \cite{SkibaSzymanska.2017}, by polarization dependent spectroscopy using a quarter wave plate and linear polarizer in front of the spectrometer. This measurement identified XX and X lines, such as for the QD with the EL spectrum in Fig.\ref{fig:DEVICE}(b) which had a fine structure splitting of (6.0$\pm$0.3)$\mu$eV.

\section*{Measurement of entanglement fidelity}
XX and X photons emitted via the biexciton cascade are co-polarized in the horizontal/vertical basis due to conservation of angular momentum \cite{Benson.2000}. The degree of correlation in a polarization basis PQ is calculated from co-polarized, c$_{PP}$, and cross-polarized, c$_{PQ}$, photon correlations by
\[
 C_{PQ}=\frac{c_{PP}-c_{PQ}}{c_{PP}+c_{PQ}}.
\]
The entanglement fidelity to the maximally entangled Bell $\phi^{+}$ state
\[
f=\frac{1+C_{HV}+C_{DA}-C_{RL}}{4},
\]
is obtained \cite{Ward.2014} from measurements in the horizontal/vertical (HV), diagonal/antidiagonal (DA), and right- and left-hand circularly polarized (RL) detection bases.

\begin{figure}[!ht]
    \centering
    \includegraphics[width=13cm,keepaspectratio]{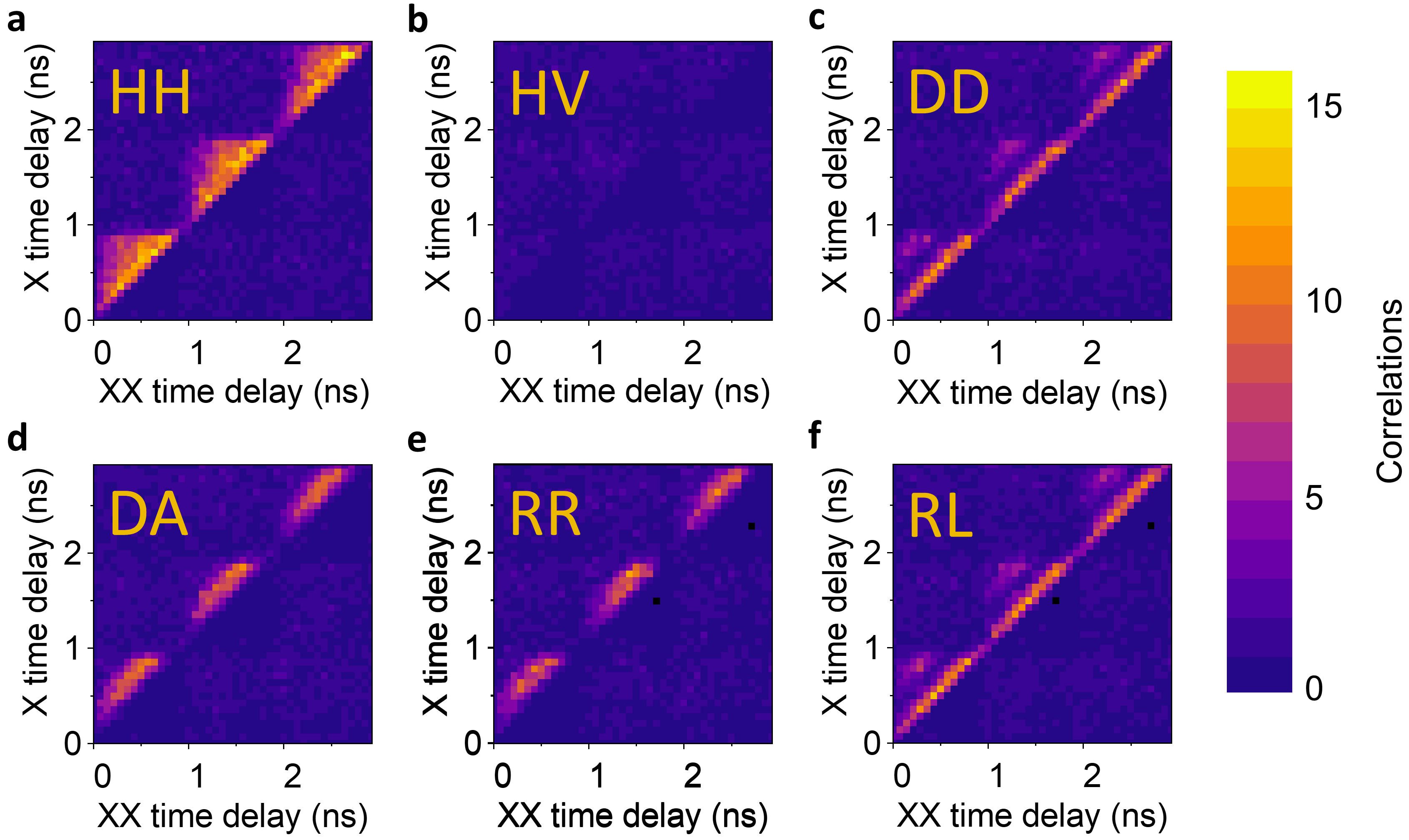}
    \caption{Normalised correlations of biexciton (XX) and exciton (X) photons as a function of XX and X delay with respect to the 1GHz clock (72ps time bins) for the laboratory measurement. The acquisition time per basis was 30 minutes. Correlations in the horizontal/vertical (HV) basis with (a) co- and (b) cross-polarised photons. Correlations in the diagonal/antidiagonal (DA) basis with (c) co- and (d) cross-polarised photons. Correlations in the right-/left-hand circular (RL) basis with (e) co- and (f) cross-polarised photons. Time bins with no photon coincidences are black.}
    \label{fig:corrLab}
\end{figure}

\begin{figure}[!ht]
    \centering
    \includegraphics[width=13cm,keepaspectratio]{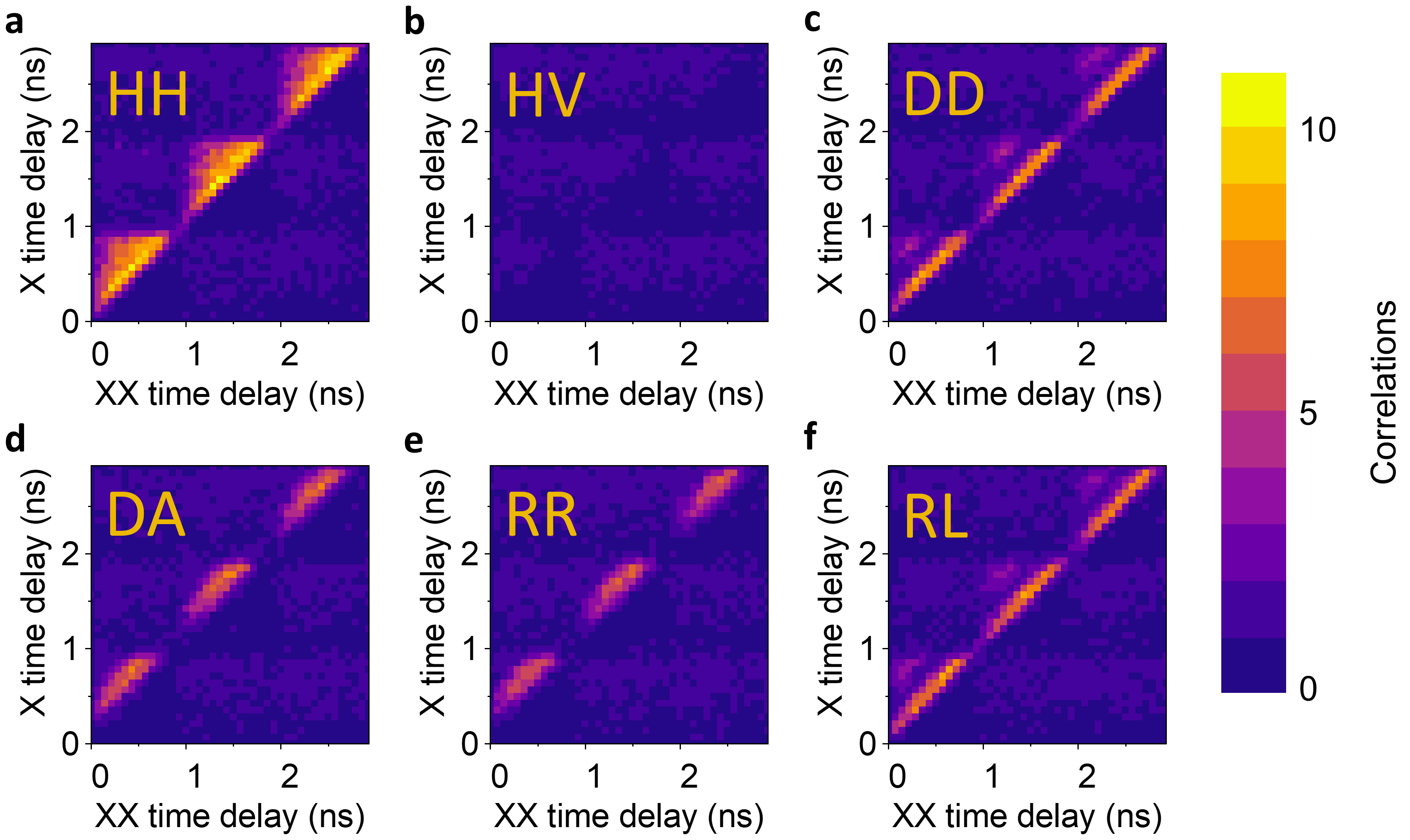}
    \caption{Normalised correlations of biexciton (XX) and exciton (X) photons as a function of XX and X delay with respect to the 1GHz clock (72ps time bins) for the entanglement distribution measurement over 15km of installed fiber. Correlations in the horizontal/vertical (HV) basis with (a) co- and (b) cross-polarised photons. Correlations in the diagonal/antidiagonal (DA) basis with (c) co- and (d) cross-polarised photons. Correlations in the right-/left-hand circular (RL) basis with (e) co- and (f) cross-polarised photons. The total acquisition time per basis was 196 minutes. In contrast to the laboratory measurement, photon coincidences for equivalent polarisation sets were combined for better statistics, such as HH and VV for (a), as all 4 polarizing beam splitter outputs were detected rather than just 3.}
    \label{fig:corrDist}
\end{figure}

Correlations in these three principal detection bases are shown for the measurement in the laboratory in Fig.\ref*{fig:corrLab} and for the entanglement distribution measurement in Fig.\ref*{fig:corrDist}. In the superposition bases, DA and RL, there is an oscillation due to the 6.0$\mu$eV fine structure splitting of the QD.

The degrees of correlation in the HV, DA, and RL bases, analysed on a 72ps grid with the reinitialisation period discarded as shown in Fig.\ref{fig:entResults}(d) and Fig.\ref{fig:distResult}(b), are C$_{HV}$= 0.87$\pm$0.05, C$_{DA}$= 0.83$\pm$0.05, and C$_{RL}$= -0.84$\pm$0.04 respectively for the laboratory measurement. Similarly, for the entanglement distribution measurement the degrees of correlation are C$_{HV}$= 0.79$\pm$0.02, C$_{DA}$= 0.68$\pm$0.02, and C$_{RL}$= -0.69$\pm$0.02.

\clearpage

A free-space spectral filter was used to separate XX and X photons in the entanglement measurements in Fig.\ref{fig:entResults}, Fig.\ref{fig:entDist}, and Fig.\ref{fig:distResult}, followed by electronic polarization controllers and polarizing beam splitters (PBS)s before the detectors.

\section*{Photon-pair coincidence rates}

\begin{figure}[!ht]
    \centering
    \includegraphics[width=9cm,keepaspectratio]{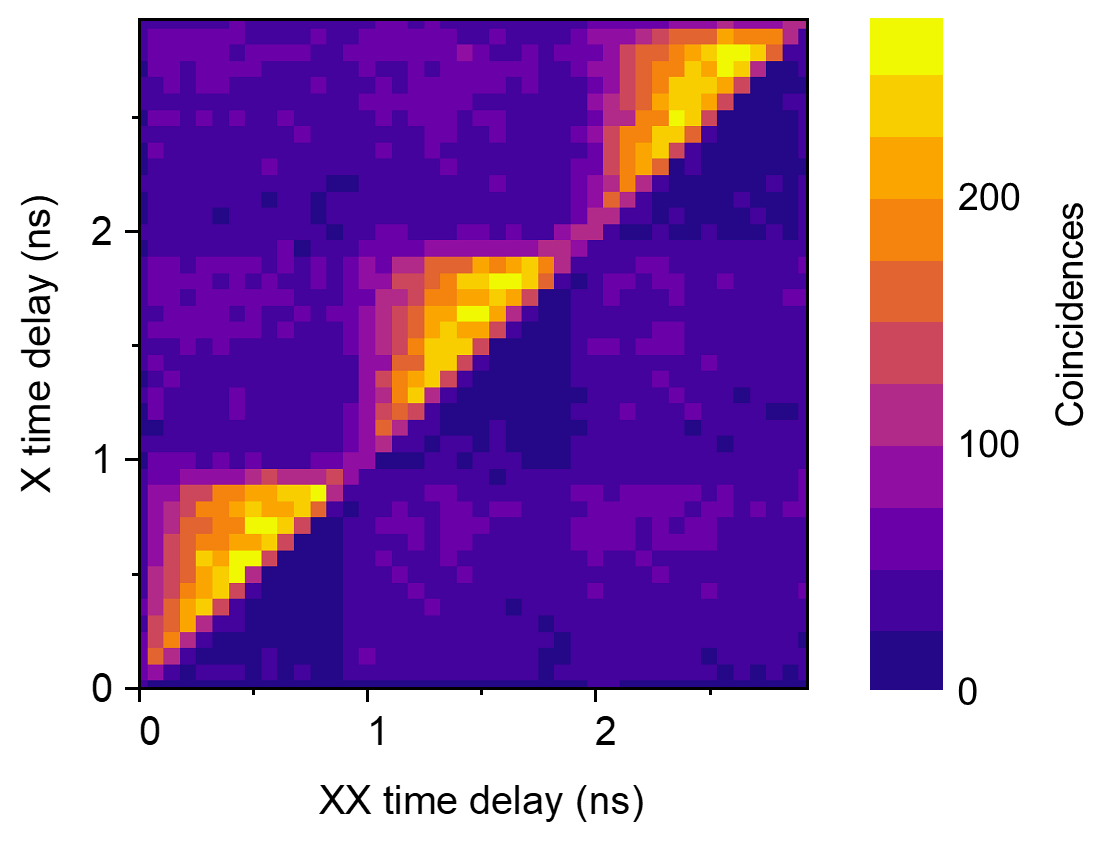}
    \caption{Total photon-pair coincidences in the 3 principal detection bases over 3 consecutive emission cycles as a function of biexciton (XX) and exciton (X) photon delays with respect to the 1GHz clock (72ps time bins).}
    \label{fig:coincs}
\end{figure}

An acquisition time of 56.25 minutes was used for the measurement of entanglement in a laboratory shown in Fig.\ref{fig:entResults}. To show the time dependence of photon-pair emission, co- and cross-polarized photon-pair coincidences in the 3 principal detection bases were summed. Fig.\ref*{fig:coincs} shows this sum with the same time bins as in Fig.\ref{fig:entResults}(b). Although the X lifetime is longer than 1ns, it can be seen that the cascade decays to the uncorrelated level within 1 excitation cycle due to the reinitialisation provided by electrical driving \cite{Muller.2020}.

52818 photon-pairs coincidences with both photons originating from the same excitation cycle were measured. However, only one output of the PBS for XX photons in Fig.\ref{fig:entResults}(a) was used for this measurement; if both outputs were used the number of coincidences would be doubled.

The overall efficiency of the optical system is approximated to be 0.6\% (22.45dB loss). The loss for coupling photons emitted by the QD into single mode fiber was ~15.2dB. The free space grating setup had a loss of 2.6dB. Typical EPC and PBS losses were 0.8dB and 0.4dB respectively. Equivalent photon loss due to finite detection efficiency of SNSPDS around 50\% was 3dB. The experimental setup had 3 fiber-to-fiber connections for XX and X photons, with losses of 0.15dB per connection.

\section*{Entanglement distribution photon detection}
At CRL, the overall timing jitter for detection was 70ps including the superconducting nanowire single photon detectors (SNSPD)s (Single Quantum). In the deployed system, the overall timing jitter for detection was around 75ps including the avalanche photodiodes (APD)s. The combined X photon rate at the SNSPDs, detectors 1 and 2 in Fig.\ref{fig:entDist}, was around 228 000 counts per second and the combined XX photon rate at the APDs, detectors 3 and 4 in Fig.\ref{fig:entDist}, was around 15 000 counts per second.

To calibrate the detection basis, the QD emission was replaced by a polarization reference matched to the eigenbasis of emitted photon pairs (not shown in Fig.\ref{fig:entDist}). EPC voltages were then varied to minimise the output signal from one output mode of the PBS at a detector, aligning the detection basis to the reference.

\clearpage
%%Bibliography
%\bibliographystyle{plainnat}
\bibliographystyle{ieeetr}
\bibliography{Refs}

\begin{thebibliography}{10}

\bibitem{Ekert.1991}
A.~K. Ekert, ``Quantum cryptography based on bell's theorem,'' {\em Physical
  review letters}, vol.~67, no.~6, p.~661, 1991.

\bibitem{Briegel.1998}
H.~J. Briegel, W.~D{\"u}r, J.~I. Cirac, and P.~Zoller, ``Quantum repeaters: the
  role of imperfect local operations in quantum communication,'' {\em Physical
  review letters}, vol.~81, no.~26, p.~5932, 1998.

\bibitem{Jacobs.2002}
B.~C. Jacobs, T.~B. Pittman, and J.~D. Franson, ``Quantum relays and noise
  suppression using linear optics,'' {\em Physical Review A}, vol.~66, no.~5,
  p.~052307, 2002.

\bibitem{Riedmatten.2004}
H.~de~Riedmatten, I.~Marcikic, W.~Tittel, H.~Zbinden, D.~Collins, and N.~Gisin,
  ``Long distance quantum teleportation in a quantum relay configuration,''
  {\em Physical review letters}, vol.~92, no.~4, p.~047904, 2004.

\bibitem{Komar.2014}
P.~Komar, E.~M. Kessler, M.~Bishof, L.~Jiang, A.~S. S{\o}rensen, J.~Ye, and
  M.~D. Lukin, ``A quantum network of clocks,'' {\em Nature Physics}, vol.~10,
  no.~8, p.~582, 2014.

\bibitem{sun.2016}
Q.~C. Sun, Y.~L. Mao, S.~J. Chen, W.~Zhang, Y.~F. Jiang, Y.~B. Zhang, W.~J.
  Zhang, S.~Miki, T.~Yamashita, H.~Terai, J.~X., C.~T. Y., Y.~L. X., C.~X. F.,
  W.~Z., F.~J. Y., Z.~Q., and P.~J. W., ``Quantum teleportation with
  independent sources and prior entanglement distribution over a network,''
  {\em Nature Photonics}, vol.~10, no.~10, pp.~671--675, 2016.

\bibitem{valivarthi.2016}
R.~Valivarthi, M.~L.~G. Puigibert, Q.~Zhou, G.~H. Aguilar, V.~B. Verma,
  F.~Marsili, M.~D. Shaw, S.~W. Nam, D.~Oblak, and W.~Tittel, ``Quantum
  teleportation across a metropolitan fibre network,'' {\em Nature Photonics},
  vol.~10, no.~10, p.~676, 2016.

\bibitem{Wengerowsky.2019}
S.~Wengerowsky, S.~K. Joshi, F.~Steinlechner, J.~R. Zichi, S.~M. Dobrovolskiy,
  R.~van~der Molen, J.~W.~N. Los, V.~Zwiller, M.~A.~M. Versteegh, A.~Mura,
  D.~Calonico, M.~Inguscio, H.~H{\"u}bel, L.~Bo, T.~Scheidl, A.~Zeilinger,
  A.~Xuereb, and R.~Ursin, ``Entanglement distribution over a 96-km-long
  submarine optical fiber,'' {\em Proceedings of the National Academy of
  Sciences}, vol.~116, no.~14, pp.~6684--6688, 2019.

\bibitem{Kwiat.1995}
P.~G. Kwiat, K.~Mattle, H.~Weinfurter, A.~Zeilinger, A.~V. Sergienko, and
  Y.~Shih, ``New high-intensity source of polarization-entangled photon
  pairs,'' {\em Physical Review Letters}, vol.~75, no.~24, p.~4337, 1995.

\bibitem{Li.2005}
X.~Li, P.~L. Voss, J.~E. Sharping, and P.~Kumar, ``Optical-fiber source of
  polarization-entangled photons in the 1550 nm telecom band,'' {\em Physical
  Review Letters}, vol.~94, no.~5, p.~053601, 2005.

\bibitem{Benson.2000}
O.~Benson, C.~Santori, M.~Pelton, and Y.~Yamamoto, ``Regulated and entangled
  photons from a single quantum dot,'' {\em Physical review letters}, vol.~84,
  no.~11, p.~2513, 2000.

\bibitem{Michler.2000}
P.~Michler, A.~Imamo{\u{g}}lu, M.~D. Mason, P.~J. Carson, G.~F. Strouse, and
  S.~K. Buratto, ``Quantum correlation among photons from a single quantum dot
  at room temperature,'' {\em Nature}, vol.~406, no.~6799, p.~968, 2000.

\bibitem{liu.2019}
J.~Liu, R.~Su, Y.~Wei, B.~Yao, S.~F.~C. da~Silva, Y.~Yu, J.~Iles-Smith,
  K.~Srinivasan, A.~Rastelli, J.~Li, and X.~Wang, ``A solid-state source of
  strongly entangled photon pairs with high brightness and
  indistinguishability,'' {\em Nature nanotechnology}, vol.~14, no.~6,
  pp.~586--593, 2019.

\bibitem{Yuan.2008}
Z.~L. Yuan, A.~R. Dixon, J.~F. Dynes, A.~W. Sharpe, and A.~J. Shields,
  ``Gigahertz quantum key distribution with ingaas avalanche photodiodes,''
  {\em Applied Physics Letters}, vol.~92, no.~20, p.~201104, 2008.

\bibitem{boaron.2018}
A.~Boaron, G.~Boso, D.~Rusca, C.~Vulliez, C.~Autebert, M.~Caloz, M.~Perrenoud,
  G.~Gras, F.~Bussi{\`e}res, M.~J. Li, D.~Nolan, A.~Martin, and H.~Zbinden,
  ``Secure quantum key distribution over 421 km of optical fiber,'' {\em
  Physical review letters}, vol.~121, no.~19, p.~190502, 2018.

\bibitem{Michler.2000b}
P.~Michler, A.~Kiraz, C.~Becher, W.~V. Schoenfeld, P.~M. Petroff, L.~Zhang,
  E.~Hu, and A.~Imamoglu, ``A quantum dot single-photon turnstile device,''
  {\em science}, vol.~290, no.~5500, pp.~2282--2285, 2000.

\bibitem{Yuan.2002}
Z.~L. Yuan, B.~E. Kardynal, R.~M. Stevenson, A.~J. Shields, C.~J. Lobo,
  K.~Cooper, N.~S. Beattie, D.~A. Ritchie, and M.~Pepper, ``Electrically driven
  single-photon source,'' {\em science}, vol.~295, no.~5552, pp.~102--105,
  2002.

\bibitem{SkibaSzymanska.2017}
J.~Skiba-Szymanska, R.~M. Stevenson, C.~Varnava, M.~Felle, J.~Huwer,
  T.~M{\"u}ller, A.~J. Bennett, J.~P. Lee, I.~Farrer, A.~B. Krysa, P.~Spencer,
  L.~E. Goff, D.~A. Ritchie, J.~Heffernan, and A.~J. Shields, ``Universal
  growth scheme for quantum dots with low fine-structure splitting at various
  emission wavelengths,'' {\em Physical Review Applied}, vol.~8, no.~1,
  p.~014013, 2017.

\bibitem{Muller.2018}
T.~M{\"u}ller, J.~Skiba-Szymanska, A.~B. Krysa, J.~Huwer, M.~Felle,
  M.~Anderson, R.~M. Stevenson, J.~Heffernan, D.~A. Ritchie, and A.~J. Shields,
  ``A quantum light-emitting diode for the standard telecom window around 1,550
  nm,'' {\em Nature communications}, vol.~9, no.~1, p.~862, 2018.

\bibitem{Santori.2001}
C.~Santori, M.~Pelton, G.~Solomon, Y.~Dale, and Y.~Yamamoto, ``Triggered single
  photons from a quantum dot,'' {\em Physical review letters}, vol.~86, no.~8,
  p.~1502, 2001.

\bibitem{Hargart.2013}
F.~Hargart, C.~A. Kessler, T.~Schwarzb{\"a}ck, E.~Koroknay, S.~Weidenfeld,
  M.~Jetter, and P.~Michler, ``Electrically driven quantum dot single-photon
  source at 2 ghz excitation repetition rate with ultra-low emission time
  jitter,'' {\em Applied Physics Letters}, vol.~102, no.~1, p.~011126, 2013.

\bibitem{Stevenson.2006}
R.~M. Stevenson, R.~J. Young, P.~Atkinson, K.~Cooper, D.~A. Ritchie, and A.~J.
  Shields, ``A semiconductor source of triggered entangled photon pairs,'' {\em
  Nature}, vol.~439, no.~7073, p.~179, 2006.

\bibitem{zhang.2015}
J.~Zhang, J.~S. Wildmann, F.~Ding, R.~Trotta, Y.~Huo, E.~Zallo, D.~Huber,
  A.~Rastelli, and O.~G. Schmidt, ``High yield and ultrafast sources of
  electrically triggered entangled-photon pairs based on strain-tunable quantum
  dots,'' {\em Nature communications}, vol.~6, p.~10067, 2015.

\bibitem{Varnava.2016}
C.~Varnava, R.~M. Stevenson, J.~Nilsson, J.~Skiba-Szymanska,
  B.~Dzur{\v{n}}{\'a}k, M.~Lucamarini, R.~V. Penty, I.~Farrer, D.~A. Ritchie,
  and A.~J. Shields, ``An entangled-led-driven quantum relay over 1 km,'' {\em
  npj Quantum Information}, vol.~2, p.~16006, 2016.

\bibitem{Muller.2020}
J.~R.~A. M{\"u}ller, R.~M. Stevenson, J.~Skiba-Szymanska, G.~Shooter, J.~Huwer,
  I.~Farrer, D.~A. Ritchie, and A.~J. Shields, ``Active reset of a radiative
  cascade for entangled-photon generation beyond the continuous-driving
  limit,'' {\em Physical Review Research}, vol.~2, no.~4, p.~043292, 2020.

\bibitem{Honjo.2007}
T.~Honjo, H.~Takesue, H.~Kamada, Y.~Nishida, O.~Tadanaga, M.~Asobe, and
  K.~Inoue, ``Long-distance distribution of time-bin entangled photon pairs
  over 100 km using frequency up-conversion detectors,'' {\em Optics express},
  vol.~15, no.~21, pp.~13957--13964, 2007.

\bibitem{Inagaki.2013}
T.~Inagaki, N.~Matsuda, O.~Tadanaga, M.~Asobe, and H.~Takesue, ``Entanglement
  distribution over 300 km of fiber,'' {\em Optics express}, vol.~21, no.~20,
  pp.~23241--23249, 2013.

\bibitem{Imamoglu.1994}
A.~Imamoglu and Y.~Yamamoto, ``Turnstile device for heralded single photons:
  Coulomb blockade of electron and hole tunneling in quantum confined p-i-n
  heterojunctions,'' {\em Physical review letters}, vol.~72, no.~2, p.~210,
  1994.

\bibitem{stevenson.2008}
R.~M. Stevenson, A.~J. Hudson, A.~J. Bennett, R.~J. Young, C.~A. Nicoll, D.~A.
  Ritchie, and A.~J. Shields, ``Evolution of entanglement between
  distinguishable light states,'' {\em Physical review letters}, vol.~101,
  no.~17, p.~170501, 2008.

\bibitem{Yuan.2007}
Z.~L. Yuan, B.~E. Kardynal, A.~W. Sharpe, and A.~J. Shields, ``High speed
  single photon detection in the near infrared,'' {\em Applied Physics
  Letters}, vol.~91, no.~4, p.~041114, 2007.

\bibitem{Chau.2002}
H.~F. Chau, ``Practical scheme to share a secret key through a quantum channel
  with a 27.6{\%} bit error rate,'' {\em Physical Review A}, vol.~66, no.~6,
  p.~060302, 2002.

\bibitem{Dousse.2010}
A.~Dousse, J.~Suffczy{\'n}ski, A.~Beveratos, O.~Krebs, A.~Lema{\^\i}tre,
  I.~Sagnes, J.~Bloch, P.~Voisin, and P.~Senellart, ``Ultrabright source of
  entangled photon pairs,'' {\em Nature}, vol.~466, no.~7303, pp.~217--220,
  2010.

\bibitem{Wang.2019}
H.~Wang, H.~Hu, T.~H. Chung, J.~Qin, X.~Yang, J.~P. Li, R.~Z. Liu, H.~S. Zhong,
  Y.~M. He, X.~Ding, Y.~H. Deng, Q.~Dai, Y.~H. Huo, S.~H{\"o}fling, C.~Y. Lu,
  and J.~W. Pan, ``On-demand semiconductor source of entangled photons which
  simultaneously has high fidelity, efficiency, and indistinguishability,''
  {\em Physical review letters}, vol.~122, no.~11, p.~113602, 2019.

\bibitem{young.2007}
R.~J. Young, S.~J. Dewhurst, R.~M. Stevenson, A.~J. Shields, P.~Atkinson,
  K.~Cooper, and D.~A. Ritchie, ``Controlling the polarization correlation of
  photon pairs from a charge-tunable quantum dot,'' {\em Applied physics
  letters}, vol.~91, no.~1, p.~011114, 2007.

\bibitem{Xiang.2020}
Z.~H. Xiang, J.~Huwer, J.~Skiba-Szymanska, R.~M. Stevenson, D.~J.~P. Ellis,
  I.~Farrer, M.~B. Ward, D.~A. Ritchie, and A.~J. Shields, ``A tuneable telecom
  wavelength entangled light emitting diode deployed in an installed fibre
  network,'' {\em Communications Physics}, vol.~3, no.~1, pp.~1--8, 2020.

\bibitem{Bockler.2008}
C.~B{\"o}ckler, S.~Reitzenstein, C.~Kistner, R.~Debusmann, A.~L{\"o}ffler,
  T.~Kida, S.~H{\"o}fling, A.~Forchel, L.~Grenouillet, J.~Claudon, and J.~M.
  G{\'e}rard, ``Electrically driven high-q quantum dot-micropillar cavities,''
  {\em Applied Physics Letters}, vol.~92, no.~9, p.~091107, 2008.

\bibitem{Ward.2014}
M.~B. Ward, M.~C. Dean, R.~M. Stevenson, A.~J. Bennett, D.~J.~P. Ellis,
  K.~Cooper, I.~Farrer, C.~A. Nicoll, D.~A. Ritchie, and A.~J. Shields,
  ``Coherent dynamics of a telecom-wavelength entangled photon source,'' {\em
  Nature communications}, vol.~5, p.~3316, 2014.

\end{thebibliography}

\end{document}